\documentclass[sigconf, nonacm]{acmart}

\newcommand{\systemicon}{\raisebox{-0.175em}{\includegraphics[height=1.0em]{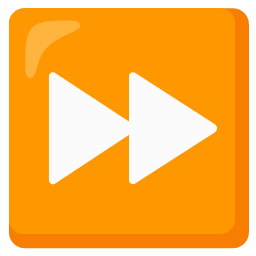}}}

\usepackage{enumitem}
\setlist[itemize]{leftmargin=25pt}

\usepackage{mathtools}

\usepackage{soul}
\usepackage{tikz}
\usepackage{listings}

\newcommand{\secref}[1]{\S\ref{#1}}

\newcommand\vldbyear{2026}
\newcommand\vldbworkshop{Applied AI for Database Systems and Applications (AIDB 2026)}
\newcommand\vldbauthors{\authors}
\newcommand\vldbtitle{\shorttitle} 

\newcommand\vldbavailabilityurl{https://github.com/mlskip/mlskip}

\newcommand\vldbpagestyle{plain} 

\usepackage{graphicx}
\usepackage{subcaption}

\newcommand{\sparagraph}[1]{\vspace{1mm}\noindent {\bf #1}}

\newcommand{\system}[0]{\textsc{MLSkip}}
\newcommand{\mlql}[0]{ML-QL}
\newcommand{\abc}[0]{$\alpha,\beta$-Crown}

\newcommand{\minmaxnumber}[0]{27.4\%}
\newcommand{\chnumber}[0]{39.3\%}
\newcommand{\bchnumber}[0]{38.31\%}

\newcommand{\bchspeedupfull}[0]{1.07$\times$}

\usepackage[dvipsnames]{xcolor}

\newcommand{\bchcolor}[1]{\colorbox{green!50!black!30}{#1}}
\newcommand{\chcolor}[1]{\colorbox{blue!20!white!70}{#1}}

\newif\ifshowcomments
\showcommentstrue

\definecolor{mscolor}{RGB}{40,120,170}     
\definecolor{mgcolor}{RGB}{150,100,40}     
\definecolor{azcolor}{RGB}{120,70,150}     
\definecolor{jvbcolor}{RGB}{0,140,120}     
\definecolor{akcolor}{RGB}{100,120,40}     
\definecolor{todocolor}{RGB}{200,0,0}      

\ifshowcomments
  \newcommand{\ms}[1]{\textcolor{mscolor}{[MS: #1]}}
  \newcommand{\mg}[1]{\textcolor{mgcolor}{[MG: #1]}}
  \newcommand{\az}[1]{\textcolor{azcolor}{[AZ: #1]}}
  \newcommand{\jvb}[1]{\textcolor{jvbcolor}{[JVB: #1]}}
  
  \newcommand{\todo}[1]{\textcolor{todocolor}{[TODO: #1]}}
\else
  \newcommand{\ms}[1]{}
  \newcommand{\mg}[1]{}
  \newcommand{\az}[1]{}
  \newcommand{\jvb}[1]{}
  \newcommand{ak}[1]{}
  \newcommand{\todo}[1]{}
\fi

\definecolor{modelcolor}{RGB}{180,0,180} 
\newcommand{\mlmodel}[1]{\textcolor{modelcolor}{\texttt{#1}}}

\definecolor{dkgreen}{rgb}{0,0.6,0}
\definecolor{gray}{rgb}{0.5,0.5,0.5}
\definecolor{mauve}{rgb}{0.58,0,0.82}
\definecolor{mygray}{rgb}{0.5, 0.5, 0.5}
\begin{document}
\title{MLSkip: Data Skipping for ML Filters via Lightweight Metadata}

\author{Mihail Stoian}
\orcid{}
\authornote{The author contributed equally to this work.}
\affiliation{%
  \institution{University of Technology Nuremberg}
  \city{Nuremberg}
  \state{Germany}
}
\email{mihail.stoian@utn.de}

\author{Mark Gerarts}
\orcid{}
\authornotemark[1]
\affiliation{%
  \institution{Hasselt University}
  \city{Hasselt}
  \state{Belgium}
}
\email{mark.gerarts@uhasselt.be}

\author{Pascal Ginter}
\orcid{}
\affiliation{%
  \institution{Technical University of Munich}
  \city{Munich}
  \state{Germany}
}
\email{pascal.ginter@tum.de}

\author{Andreas Zimmerer}
\orcid{0000-0002-4158-5805}
\affiliation{%
  \institution{University of Technology Nuremberg}
  \city{Nuremberg}
  \state{Germany}
}
\email{andreas.zimmerer@utn.de}

\author{Jan Van den Bussche}
\orcid{}
\affiliation{%
  \institution{Hasselt University}
  \city{Hasselt}
  \state{Belgium}
}
\email{jan.vandenbussche@uhasselt.be}

\author{Andreas Kipf}
\orcid{}
\affiliation{%
  \institution{University of Technology Nuremberg}
  \city{Nuremberg}
  \state{Germany}
}
\email{andreas.kipf@utn.de}

\begin{abstract}
Database vendors recently released AI functions that can be used in filter predicates. As such functions often rely on costly, black-box ML models, they unveil new data management challenges. Concretely, traditional data skipping techniques for integer and string data fail to be applicable to the new filter type.
Indeed, there is no known mechanism for pruning non-qualifying row groups, e.g., when reading files from blob storage.

In this work, we initiate the study of data skipping techniques for ML filters. We make the case that Parquet's default min-max metadata is enough to enable pruning.
To this end, we draw connections to two lines of research: (i) the recently proposed query language for ML models and (ii) neural network verification.

Our preliminary results on ReLU architectures show that on tables from TPC-H and TPC-DS, the average pruning effectiveness for filters of selectivity below 0.1\% amounts to \minmaxnumber{}.
Finally, inspired by research on spatial joins, we propose an enhanced metadata structure: a size-bounded 2D convex hull that verification tools can make better use of, increasing the pruning effectiveness to \bchnumber{}, while occupying at most 45 bytes per row group and column pair. We observe an end-to-end speedup of \bchspeedupfull{} over PyTorch in DuckDB.
\end{abstract}

\maketitle

\pagestyle{\vldbpagestyle}
\begingroup\small\noindent\raggedright\textbf{VLDB Workshop Reference Format:}\\
\vldbauthors. \vldbtitle. VLDB \vldbyear\ Workshop: \vldbworkshop.\\
\endgroup
\begingroup
\renewcommand\thefootnote{}\footnote{\noindent
This work is licensed under the Creative Commons BY-NC-ND 4.0 International License. Visit \url{https://creativecommons.org/licenses/by-nc-nd/4.0/} to view a copy of this license. For any use beyond those covered by this license, obtain permission by emailing \href{mailto:info@vldb.org}{info@vldb.org}. Copyright is held by the owner/author(s). Publication rights licensed to the VLDB Endowment. \\
\raggedright Proceedings of the VLDB Endowment.
ISSN 2150-8097. \\
}\addtocounter{footnote}{-1}\endgroup

\ifdefempty{\vldbavailabilityurl}{}{
\vspace{.3cm}
\begingroup\small\noindent\raggedright\textbf{VLDB Workshop Artifact Availability:}\\
The source code, data, and/or other artifacts have been made available at \url{\vldbavailabilityurl}.
\endgroup
}

\section{Introduction}\label{sec:introduction}

Support for executing ML models over relational data has become commonplace in modern database systems; examples include Google BigQuery's \texttt{ML.PREDICT}~\cite{bigquery-mlpredict}, Databricks' \texttt{ai\_query}~\cite{databricks-ai_query}, Microsoft T-SQL's \texttt{PREDICT}~\cite{microsoft-predict}, Snowflake's Model Registry~\cite{snowflake-mr}, Amazon Redshift ML~\cite{redshift-ml}, Oracle's \texttt{OML4SQL}~\cite{oracle-oml4sql}, and many others.

\sparagraph{ML Filters.} Indeed, integrating ML inference into the query engine increases flexibility and encourages customers to keep both their data and ML workloads within the database system~\cite{raven-mlsql-2}. The following example shows how a neural network, {\small\mlmodel{sanity\_score\_nn}}, can be used as a scoring function inside a table scan:\footnote{The model evaluates the ``sanity score'' of a city; see Ref.~\cite{demographia} for a formal analysis.}

\begin{lstlisting}[label=lst:city-sanity,language=SQL]
SELECT city_name FROM cities WHERE
  (*@\mlmodel{sanity\_score\_nn}@*)(population_density, housing_cost) > 0.9;
\end{lstlisting}
As the complexity of the filter increases, the query engine loses the ability to effectively prune non-qualifying row groups, a capability that major vendors optimize for traditional filters~\cite{snowflake-pruning}.

\sparagraph{Research Question.} Unlike traditional equality and range table predicates, ML filters are difficult for query engines to reason about. Traditional predicates benefit from metadata-aware file formats such as Parquet, which maintain min-max statistics that enable data skipping. For example, a predicate \texttt{housing\_cost > 3k} can exclude a row group with (\texttt{min}, \texttt{max}) = (1k, 2k). Given the recent adoption of ML filters, it is natural to ask whether it is possible to enable data skipping by leveraging the built-in metadata, available on both managed and open file formats.

\sparagraph{\systemicon{} \system{}.} Somewhat surprisingly, we show in this work that, \emph{even with the already available min-max metadata}, one can obtain measurable pruning effects for standard ML models such as ReLU architectures.
To this end, we draw connections to two lines of research, namely: (i) the recently proposed query language for ML models~\cite{grohe2024query}, and (ii) the established line of research on neural network verification, featuring tools such as Marabou~\cite{marabou1} and \abc{}~\cite{Kotha2023ProvablyBoundingNeural}. Our contributions are as follows:
\begin{itemize}
    \item[(1)] We introduce \system{}, the first framework for row group pruning for ML filters.
    \item[(2)] We evaluate the effectiveness of the two approaches with min-max metadata on the TPC-H~\cite{tpch} and TPC-DS~\cite{tpcds} benchmarks (sf = 1), using ReLU architectures trained on 2-4 features (columns), and obtain an average pruning effectiveness of \minmaxnumber{} for filters of $\leq 0.1\%$-selectivity.\footnote{pruning effectiveness [$\%$] = \#\texttt{pruned\_row\_groups} / \#\texttt{prunable\_row\_groups}.}
    \item[(3)] We propose enhanced, geometry-aware metadata that boosts the pruning effectiveness to \bchnumber{}, reaching an end-to-end speedup of \bchspeedupfull{} over PyTorch in DuckDB across all filters.
\end{itemize}

\sparagraph{Related Work.} Integrating ML inference in a query engine is a rather recent idea, with origins in many proposals in the last decade~\cite{raven-mlsql-2, smart-mlsql-3, aero-mlsql-1}.
After the arrival of LLMs, the simple ML operators have been replaced by the more flexible semantic operators~\cite{aggarwal2025cortex, google-proxy-models, 2025-vldb-docetl, patel2024lotus, liu2025palimpzest, jo2024thalamusdb, dorbani2025flockmtl}.
Notably, recent work on optimized semantic filters showed that traditional ML models, such as logistic regression, can replace expensive LLMs for query processing for certain cases~\cite{google-proxy-models}. The idea is to use some initial sample of the table---which is processed with an expensive LLM---as training data for a proxy model over pre-computed row-embeddings (in this case a logistic regression model, but other models might be used as well).
After the proxy model has been trained, its accuracy is compared against the LLM and, if deemed good enough, execution is continued with the proxy model instead of the LLM.
We can imagine that \system{} can be used for such optimizations of semantic filters via proxy models to further speed up query processing by allowing to skip evaluation of entire row groups.

Research in the context of traditional ML-native engines includes Raven~\cite{raven-mlsql-2}, Smart~\cite{smart-mlsql-3}, Umbra's PyTorch extension~\cite{riegger-umbra}, and LingoDB's MLIR-empowered model compilation~\cite{lingodb}.
However, despite much research on this topic, we are unaware of any work on data skipping techniques for this new filter type.
\section{Preliminaries}\label{sec:preliminaries}

In this section, we briefly present file format metadata such as that of Parquet's~\secref{subsec:metadata}, neural network verification~\secref{subsec:nnv}, and the recent paradigm of query languages for ML models~\secref{subsec:mlql}.

\subsection{File Format Metadata}\label{subsec:metadata}

The quest for data skipping led modern file formats like Parquet to integrate metadata column statistics, including the min-max values of the column, the number of \texttt{NULL} values, and the distinct count~\cite{parquet-metadata}; these are missing in traditional file formats like CSV.
Given Parquet's widespread adoption, query engines highly optimized for this (limited) metadata structure~\cite{snowflake-pruning}.
Noteworthy, Parquet supports adding custom metadata, such as our enhanced 2D metadata (\secref{subsec:enhanced-metadata}), while preserving the readability compatibility~\cite{parquet-blogpost}.


\subsection{Neural Network Verification}\label{subsec:nnv}

Neural networks (NNs) are indispensable tools in today's software landscape. When used 
in mission-critical systems, one must have guarantees on how the networks 
operate. The field of neural network verification (NNV) develops tools to
formally verify properties such as robustness, consistency, and 
reachability~\cite{albarghouthi2021introduction,Liu2021AlgorithmsVerifyingDeep}.
We focus on the latter: given a neural network and some input constraints, we want to 
verify if the network's output maps to a given value. 
 
There are several tools to verify NNs. In this work, we consider 
Marabou~\cite{marabou1, marabou2}, due to its easy setup and fast verification time. The state-of-the-art tool is \abc{}~\cite{Kaulen2025vnncomp,Kotha2023ProvablyBoundingNeural},
which scales to larger models by sacrificing completeness for
efficiency~\cite{Wang2021BetaCROWNEfficient},\footnote{That is, a property proven correct
by \abc~is always correct, but a property that
\abc~fails to verify might still be correct in reality. In the
context of data skipping, this is acceptable: we may read some row groups that could have been pruned, but never miss any that are required.} and has a GPU-optimized implementation, making it compelling for the rise of GPU-enabled databases~\cite{tqp, bowen-gpu}.

\subsection{Query Languages for ML Models}\label{subsec:mlql}

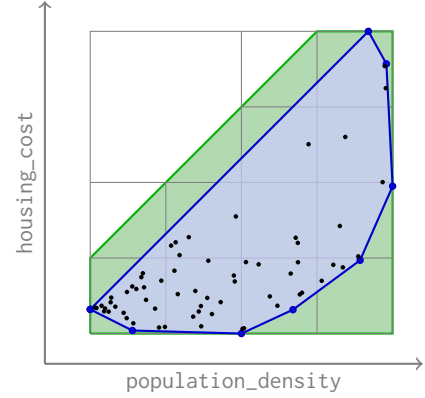
\begin{figure}[t]
\centering
\begin{tikzpicture}[scale = 2]
  
  \draw[green!70!black, fill=green!50!black, fill opacity=0.3, thick] plot[mark=square*, mark size=0.75pt] coordinates {
    (-1, -1) (-1, -0.5) (0.5, 1) (1, 1) (1, -1) (-1, -1)
  };


\draw[gray, thick, ->] (-1.3, -1.2) -- (1.2, -1.2);
\draw[gray, thick, ->] (-1.3, -1.2) -- (-1.3, 1.2);

\node[anchor=north] at (-0.05, -1.2)
    {\textcolor{gray}{\texttt{population\_density}}};

\node[rotate=90, anchor=south] at (-1.3, 0)
    {\textcolor{gray}{\texttt{housing\_cost}}};

  \draw[gray] (-1, -1) -- (-1, 1) -- (1, 1) -- (1, -1) -- (-1, -1);
  \draw[gray] (0, -1) -- (0, 1);
  \draw[gray] (-1, 0) -- (1, 0);

  \draw[gray] (-1, -0.5) -- (0, -0.5);
  \draw[gray] (-0.5, -1) -- (-0.5, 0);

  \draw[gray] (0, -0.5) -- (1, -0.5);
  \draw[gray] (0.5, -1) -- (0.5, 0);

  \draw[gray] (0, 0.5) -- (1, 0.5);
  \draw[gray] (0.5, 1) -- (0.5, 0);

  \draw[blue!80!black, fill=blue!20, fill opacity=0.7,thick] plot[mark=*, mark size=0.5pt, mark options={blue!80!black}] coordinates {
    (-1.0,-0.84) (-0.721,-0.98) (0.0,-1.0) (0.341,-0.842)
    (0.785,-0.515) (1.0,-0.025) (0.959,0.785) (0.84,1.0) (-1.0,-0.84)
  };

  \foreach \p in {
    (-0.05,-0.616), (0.326,-0.603), (-0.959,-0.831), (-0.443,-0.584), (0.374,-0.528),
    (0.114,-0.542), (-0.695,-0.705), (-0.662,-0.627), (-0.865,-0.793), (-0.834,-0.822),
    (-0.319,-0.888), (0.671,-0.562), (-0.861,-0.762), (0.017,-0.965), (0.955,0.624),
    (-0.759,-0.898), (-0.715,-0.932), (0.238,-0.816), (0.386,-0.741), (-0.644,-0.694),
    (-0.195,-0.871), (0.188,-0.754), (-0.226,-0.764), (0.444,0.253), (0.934,0.002),
    (-0.282,-0.854), (0.373,-0.4), (-0.757,-0.725), (-0.268,-0.818), (0.608,-0.545),
    (-0.925,-0.816), (-0.137,-0.792), (-0.545,-0.96), (-0.219,-0.518), (-0.779,-0.863),
    (-0.631,-0.777), (-0.908,-0.853), (-0.305,-0.718), (-0.878,-0.855), (-0.571,-0.835),
    (-0.037,-0.225), (-0.652,-0.601), (-0.266,-0.952), (0.687,0.301),
    (-0.967,-0.827), (0.531,-0.651), (-0.506,-0.956), (-0.902,-0.834), (0.948,0.77),
    (0.007,-0.969), (-0.348,-0.362), (0.651,-0.288), (-0.53,-0.65), (-0.044,-0.655),
    (-0.465,-0.418), (0.773,-0.489), (-0.409,-0.481), (-0.435,-0.396), (0.03,-0.527),
    (0.401,-0.73), (-0.421,-0.739), (0.359,-0.366), (-0.722,-0.688)
  }{ \fill[black] \p circle (0.015); }

\end{tikzpicture}
\caption{Row group metadata variants: (i) vanilla min-max ranges induce a 2D rectangle, (ii) \chcolor{\texttt{ConvexHull}} creates the optimal convex polygon around the row group's data points, while (iii) \bchcolor{\texttt{BoundedConvexHull}} bounds the number of vertices in the hull, by working on a grid of depth 2.}
\label{fig:metadata-vis}
\end{figure}

Recent work has shown that SQL can act as a query language for neural
networks~\cite{grohe2024query}. By representing a network's structure and
weights as a database instance, SQL can reason about the network and verify its
behavior. We implemented this paradigm
as a two-step approach, which we call \mlql\@. First, \mlql~compiles the network to a
geometric representation and stores it in a database. We then use this 
precomputed representation to execute many
verification queries in SQL relatively cheaply, unlike traditional tools 
that run entirely at query time~\cite{katz2017reluplex}.

\begin{figure*}[t]
    \centering
    \begin{subfigure}[t]{0.49\textwidth}
        \centering
        \includegraphics[width=\linewidth]{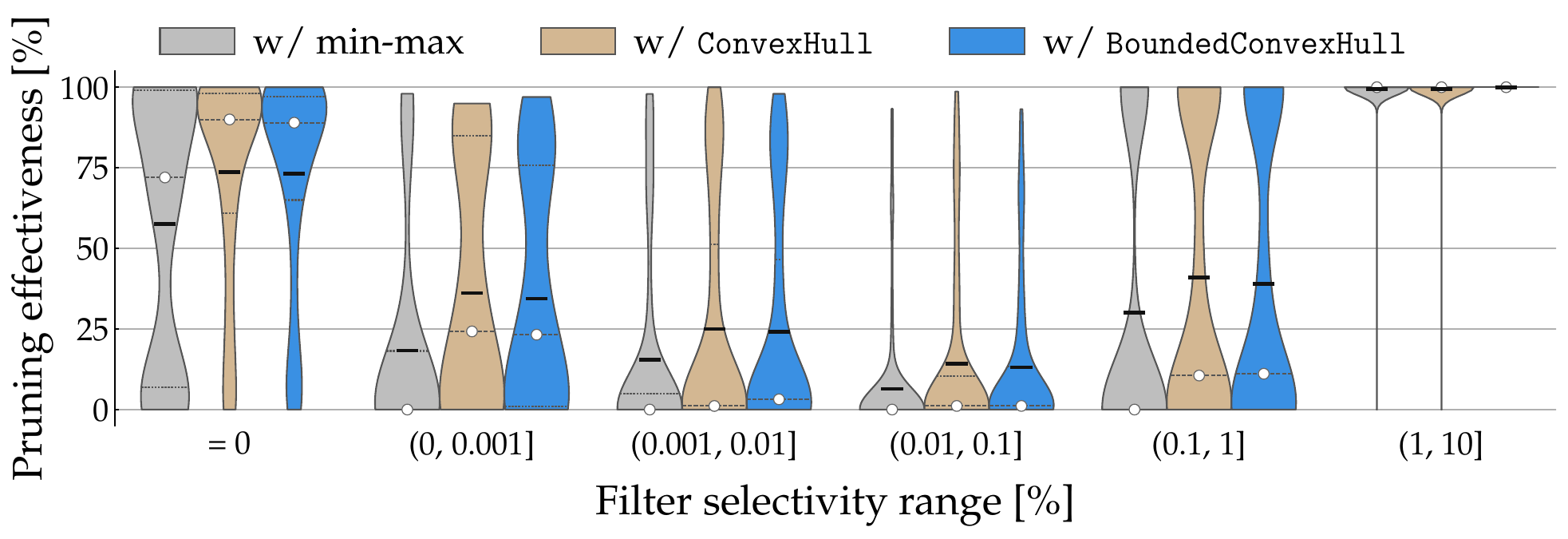}
        \caption{Row group size = 1K}
        \label{fig:sub1}
    \end{subfigure}
    \hfill
    \begin{subfigure}[t]{0.49\textwidth}
        \centering
        \includegraphics[width=\linewidth]{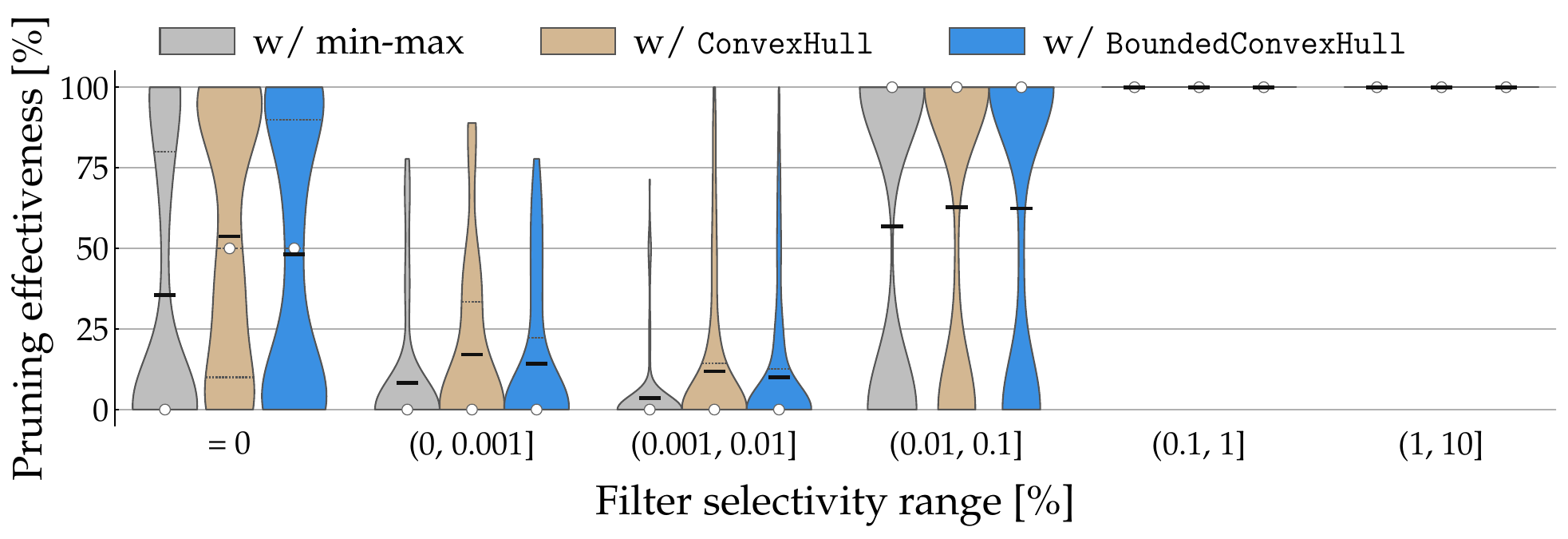}
        \caption{Row group size = 10K}
        \label{fig:sub2}
    \end{subfigure}
    \caption{Data skipping behavior for varying filter selectivity under two row group sizes $\in \{\text{1K}, \text{10K}\}$ and the metadata types. The models are neural networks with 2 hidden layers trained on 2-4 table columns of the TPC-H and TPC-DS benchmarks.}
    \label{fig:pruning-plot}
\end{figure*}

We consider feed-forward networks with ReLU activation functions and a single
output node. Such networks represent piecewise linear functions.
Geometrically, the ReLU layers partition the continuous input space into
a hyperplane arrangement: a collection of polyhedral regions where the network behaves
as an affine function within each distinct region.

Prior work reconstructs the output functions of networks with a 
single input node and a single hidden layer~\cite{Gerarts2025SQL4NNValidationexpressive}; we generalize 
this approach to networks with an arbitrary number of inputs and layers. Our compilation
step is based on
Cylindrical Algebraic Decomposition (CAD), mirroring the proof of the theoretical foundations~\cite{grohe2024query}.
However, CAD imposes restrictions on the size of networks we can practically
verify because it scales doubly exponential~\cite{Collins1975Quantifiereliminationreal} in the number 
of input features of the network, and the number of inputs
for the algorithm grows exponentially in the number of hidden layers due to ReLU activation functions. 
Future work is needed to replace CAD with more efficient algorithms.

We use the geometric representation to answer reachability
queries: determining whether the network's output can fall within
a specific range given a set of input constraints. While many alternative 
techniques exist for this particular 
problem~\cite{Xiang2017ReachableSetComputation,Vincent2025ReachablePolyhedralMarching}, 
viewing the problem through a database lens provides a distinct advantage: it shifts the heavy 
lifting entirely to the offline phase, allowing the online verification phase to benefit directly from 
database query optimization, e.g., we can batch all table metadata in the same SQL query. The verification queries currently use approximation techniques and are 
thus incomplete, just like \abc. This is a practical constraint; in future work, we plan to use exact techniques.

To compare against other verifiers, we evaluate \mlql~as a standalone tool.
However, we envision that a sufficiently advanced query planner might use the
geometric representation directly to optimize data skipping. This remains a topic for
future work.

\section{\system{}}\label{sec:approach}

\system{} is the first framework to enable data skipping for ML filters, using lightweight metadata, such as Parquet's.  \system{} works as follows. Consider the query used in the introduction. Assuming already available min-max metadata for each of the two columns, we specify the input space as the rectangle spanned by the min-max ranges, namely: $[\texttt{min1}, \texttt{max1}] \times [\texttt{min2}, \texttt{max2}]$.\footnote{In our example: \texttt{min1} $\vcentcolon=$ \texttt{select min(population\_density) from cities}.}
Next, we specify the input constraints, along with the output constraint $(0.9, +\infty)$ from the filter, i.e., {\small \mlmodel{sanity\_score\_nn}\texttt{(...) > 0.9}}, and obtain the result whether the NN's function, on that input, takes values in that output constraints.
If the result is negative, we can safely skip the respective row group.

In this work, our goal is to study both (i) the verification tools (NNV, ML-QL) and (ii) the metadata required to enable the skipping.
Next, we detail on the latter point.

\subsection{Enhanced Metadata}\label{subsec:enhanced-metadata}

Observing the poor, yet non-zero pruning performance in the presence of only min-max metadata, we propose in the following a more enhanced metadata design, inspired from research on spatial joins~\cite{spatial-joins-seeger} and linear optimization queries~\cite{onion}.
We gradually build towards a compact metadata, that, interestingly, fits well with the input structure expected by the verifiers.
We first consider the case of a 2-feature model, as per our sample model {\small\mlmodel{sanity\_score\_nn}}.
The following ideas are visualized in Fig.~\ref{fig:metadata-vis}.

\sparagraph{\chcolor{\texttt{ConvexHull}}.} The first idea is to build a convex hull of the two-dimensional data points.
The intuition is that regions around the corners induced by the 2D min-max ranges are unoccupied, leading to a smaller input domain.
However, the disadvantage of vanilla convex hulls is that we cannot bound the number of stored vertices, leading to unpredictable metadata size (and verification time).

\sparagraph{\bchcolor{\texttt{BoundedConvexHull}}.} The second idea, which does guarantee bounded metadata size, is to instead recursively split the 2D rectangle induced by the min-max ranges into a grid of given depth; unoccupied grid cells are not split.
As performing verification on the grid cells would be particularly expensive,\footnote{Indeed, one would have to perform a verification call for each occupied grid cell.} we instead fit a convex hull on top of the occupied cells.
This comes with two clear advantages: (i) the verification becomes single-pass, as we simply constraint the input with the linear constraints of the convex hull,\footnote{This property also holds for min-max metadata, allowing for fast verification time.} and (ii) the metadata size is bounded, as the number of hull vertices is, by grid's construction, bounded. We store a $2d + 2$ bit per vertex, capturing the cell's index ($2 \times d$ bit) and the involved corner (2 bit).

\sparagraph{Integration.} Our 2D metadata can also be used when using $d$-dimensional models, without requiring the presence of metadata on more than two dimensions.
Namely, we will constraint the input space via all ${d \choose 2}$ (bounded) convex hulls.
Currently, we build the 2D metadata on all column pairs. In a production setting, one should analyze the workload to understand which pairs are useful for pruning so that the build overhead pays off.
\section{Evaluation}\label{sec:evaluation}

\sparagraph{Setup.} To train the NNs, we use the first 2k table rows of the TPC-H and TPC-DS benchmarks (sf = 1).
The functions are learning a column using 2-4 other table columns;
in total, we have 10 regression function templates, proposed by OpenAI's GPT5.5 model: 2 on TPC-H and 8 on TPC-DS. The trained NNs have either 1 or 2 hidden layers, each with 32 nodes;
hence, the largest neural network has 1317 parameters.
We generate the filters with varying selectivity, by gradually increasing the width of the queried filter range and picking a random range left bound, following the structure:
\begin{lstlisting}[label=lst:city-sanity,language=SQL]
    where model(col1, ..., coln) between a and b;
\end{lstlisting}
We obtain a total of 1376 such filters of varying selectivity. The \texttt{BoundedConvexHull} metadata comes with a default grid depth of 4. We conduct the experiments on a single node Intel\textsuperscript{\textregistered} Xeon\textsuperscript{\textregistered} Gold 5318Y CPU (24 cores, 48 hyper-threads). The machine is equipped with 128GB DDR4 main memory and runs Ubuntu 24.04.

\sparagraph{Tools.} We use Marabou v2.0~\cite{marabou2} and our implementation of ML-QL for neural networks for multiple inputs (\secref{subsec:mlql}). ML-QL is currently limited to models with a single hidden layer. Due to its slow model compilation phase, we only benchmark it on two-dimensional models; it takes around 0.8s to compile the model with CAD, stored then in 3 tables, the largest having 32k rows and 7 columns.

\subsection{Pruning Effectiveness}

We report the \emph{pruning effectiveness} of Marabou for all trained ML filters, computed as: \[
  \frac{\#\texttt{pruned\_row\_groups}}{\#\texttt{prunable\_row\_groups}} \cdot 100.
\]
In Fig.~\ref{fig:pruning-plot}, we report the pruning effectiveness achieved across both benchmarks for row groups of 1K and 10K rows, clustering the filters in selectivity buckets. (Note that highly non-selective filters will also have a high effectiveness by default.) We observe that for zero-selectivity filters, the pruning is quite effective, especially when enabling the enhanced metadata (\secref{subsec:enhanced-metadata}).
For highly selective filters ($\leq$ 0.1\% selectivity), with min-max metadata, we obtain an average effectiveness of \minmaxnumber{}. Notably, \texttt{BoundedConvexHull} metadata improves this number to \bchnumber{}, while \texttt{ConvexHull} gets to \chnumber{}.
Naturally, when running on groups consisting of 10K rows, the pruning effect is reduced on the same selectivity buckets, as the information captured by the metadata becomes coarser.

\begin{figure}[t]
    \centering
    \includegraphics[width=1.0\linewidth]{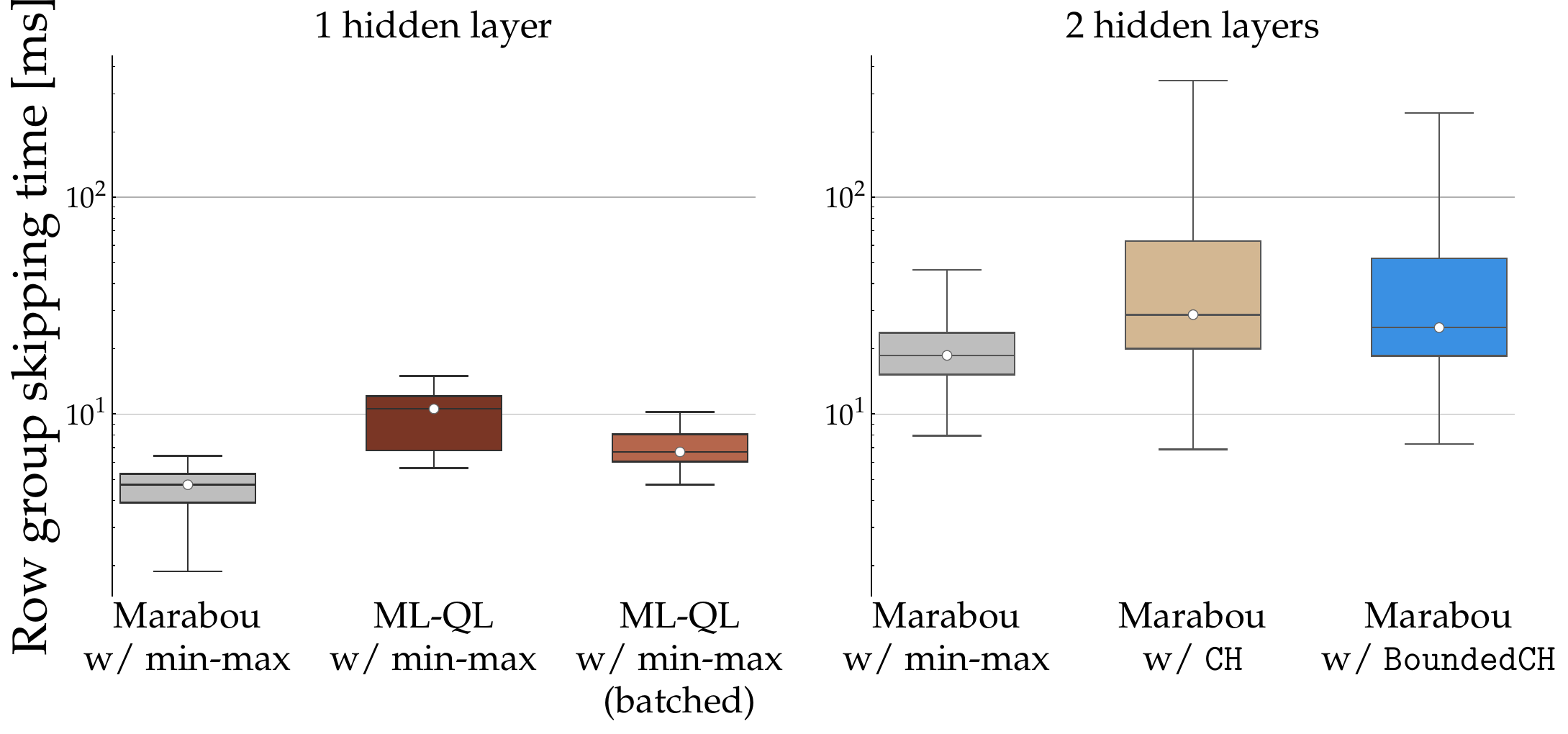}
    \caption{Data skipping time for models on TPC-H and TPC-DS with a row group size of 1K.}
    \label{fig:block-skipping-time}
\end{figure}

\begin{figure}[t]
    \centering
    \includegraphics[width=1.0\linewidth]{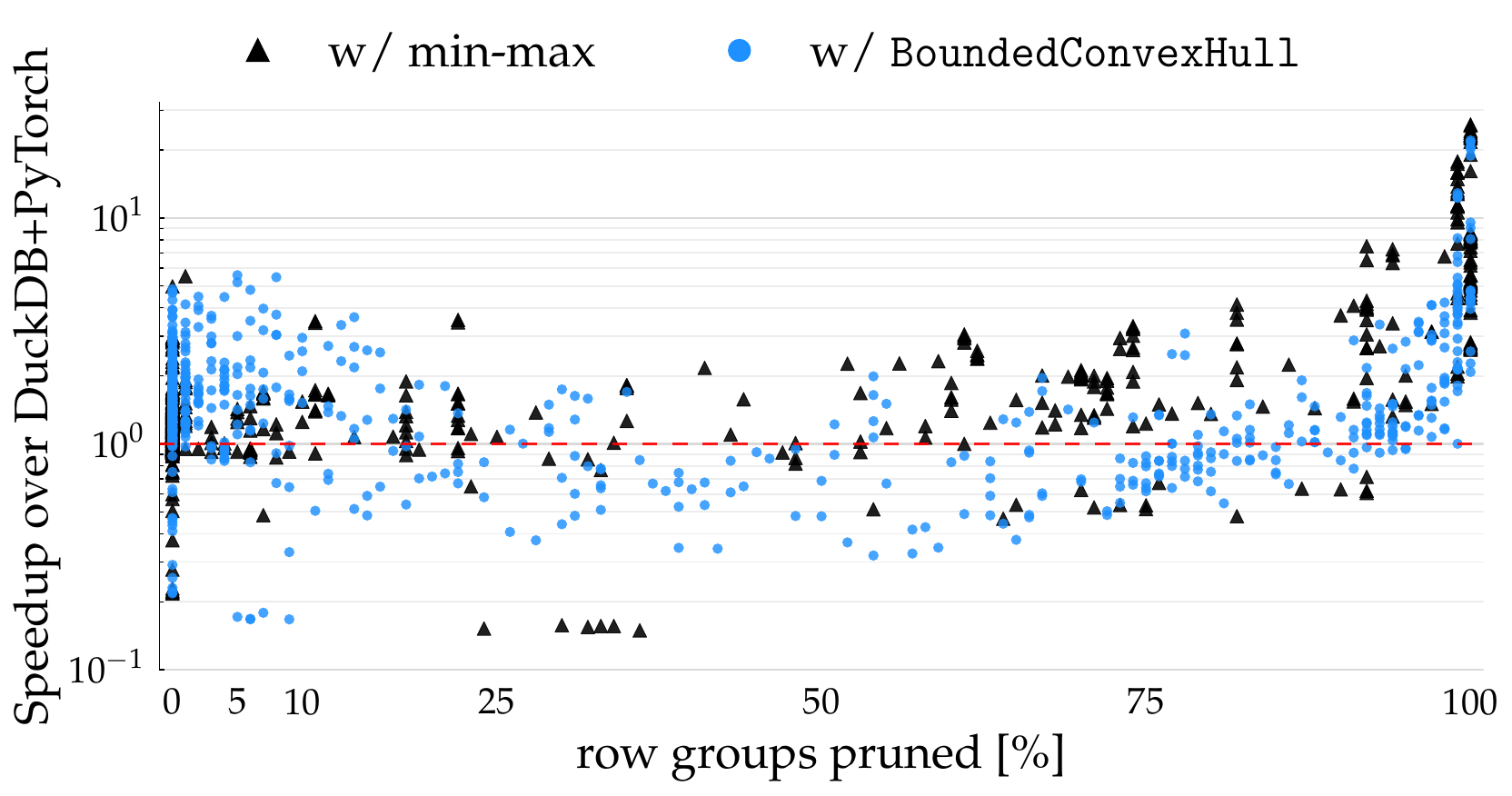}
    \caption{End-to-end speedup on DuckDB with PyTorch across all benchmark filters, using a row group size of 1K.}
    \label{fig:e2e-speedup}
\end{figure}

\sparagraph{E2E Speedup.} Being able to prune row groups makes model inference cheaper. Indeed, when running PyTorch 2.8.0+cpu~\cite{pytorch} in DuckDB 1.5~\cite{duckdb}, we observe a total end-to-end speedup of \bchspeedupfull{} across \emph{all} filters, when using models with 2 hidden layers. We plot in Fig.~\ref{fig:e2e-speedup} the relationship between the pruning ratio and the speedup. 
Where MLSkip is unable to prune a substantial fraction of row groups, e.g., $\leq$ 5\%, the (noisy) overhead of verification tends to dominate.

\subsection{Pruning Time}

We report the time to prune a row group via both Marabou and ML-QL. We differentiate between using models with 1 and 2 hidden layers, respectively.
First, ML-QL is around 2x slower than Marabou on models with one hidden layer, with our batched variant narrowing this gap.
Second, the time increases by almost one magnitude for Marabou with min-max metadata on models with two hidden layers.
Noteworthy, the effect of having a bounded number of vertices in the convex hull can be observed for \texttt{BoundedConvexHull}, inducing a faster verification time than with \texttt{ConvexHull}.

\subsection{Metadata Overhead}

While building min-max metadata is rather cheap, our enhanced metadata requires more compute (and more space). We show this overhead in terms of both compute and space in Tab.~\ref{tab:metadata-overhead}, for table column pairs across 100 row groups. Naturally, min-max metadata requires $2 \times (8 + 8) = 32$ byte to store a pair of doubles per column; the build time is that of a DuckDB table scan.

The regular \texttt{ConvexHull} requires two doubles for each convex hull vertex. Hence, without having control on the number of vertices, we observe that the overhead can get up to 304 byte, making this metadata type rather impractical; note that the lightweight HyperLogLog sketch~\cite{hyperloglog} is usually optimized to occupy 64 byte~\cite{freitag-groupby}. Our \texttt{BoundedConvexHull} achieves this goal, by fitting the convex hull on a grid support. In this case, we empirically observed a maximum size of 45 byte (for a grid depth of 4). Note that the space footprint can be reduced by choosing a smaller grid depth. Both metadata types are built via \texttt{scipy.spatial.ConvexHull}~\cite{scipy-convexhull}. To support a $d$-dimensional input, the metadata has to be built on all $d \choose 2$ pairs; see~\secref{subsec:enhanced-metadata} for a discussion on how to optimize this step.

\begin{table}[h]
\caption{Metadata space footprint and build time per row group for column pairs; the default grid depth is 4.}
\label{tab:metadata-overhead}
\centering
\small
\begin{tabular}{l|cc|cc}
\toprule
Metadata type & \multicolumn{2}{c}{Size [byte]} & \multicolumn{2}{c}{Build time [ms]} \\
 & avg & max & avg & max \\
\midrule
 min--max & 32.00 & 32.00 & 12.09 & 20.09 \\
 \texttt{ConvexHull} & 161.12 & 304.00 & 31.73 & 41.22 \\
 \texttt{BoundedConvexHull} & 40.51 & 45.00 & 31.74 & 41.59 \\
\bottomrule
\end{tabular}
\end{table}

\section{Conclusion \& Future Work}\label{sec:conclusion}

With \systemicon{} \system{}, we showed that it is indeed possible to enable pruning for the rather complex ML filters, with focus on ReLU architectures, even when using only lightweight metadata, while richer metadata can further improve pruning.
Given the rise of semantic query processing and its use of proxy models~\cite{google-proxy-models}, \system{} is a necessary tool to enable faster query times and save I/O from blob storage. We plan to test MLSkip with larger models, e.g., Transformers~\cite{transformer}, study verification on string data, and integrate a Batch API into current neural network verification tools.


\bibliographystyle{ACM-Reference-Format}
\bibliography{mlskip}

\end{document}